\documentclass{article}
\usepackage[preprint]{neurips_2022}
\usepackage{amsmath,amsfonts}
\usepackage[ruled]{algorithm2e}
\usepackage{array}
\usepackage[caption=false,font=normalsize,labelfont=sf,textfont=sf]{subfig}
\usepackage{textcomp}
\usepackage{stfloats}
\usepackage{url}
\usepackage{verbatim}
\usepackage{graphicx}
\usepackage{bm}
\usepackage{booktabs}
\usepackage{multirow}

\begin{document}
	
\title{Low-Dose CT Using Denoising Diffusion Probabilistic Model for 20$\times$ Speedup}

\author{Wenjun Xia, Qing Lyu, Ge Wang\\
	\normalfont Department of Biomedical Engineering \\
	Rensselaer Polytechnic Institute\\
	Troy, NY 12180 USA\\
	\texttt{\{xiaw4, lyuq, wangg6\}@rpi.edu}
}



\maketitle

\begin{abstract}
	Low-dose computed tomography (LDCT) is an important topic in the field of radiology over the past decades. LDCT reduces ionizing radiation-induced patient health risks but it also results in a low signal-to-noise ratio (SNR) and a potential compromise in the diagnostic performance. In this paper, to improve the LDCT denoising performance, we introduce the conditional denoising diffusion probabilistic model (DDPM) and show encouraging results with a high computational efficiency. Specifically, given the high sampling cost of the original DDPM model, we adapt the fast ordinary differential equation (ODE) solver for a much-improved sampling efficiency. The experiments show that the accelerated DDPM can achieve 20x speedup without compromising image quality.
\end{abstract}

\section{Introduction}
	Computed tomography (CT) plays an indispensable role in radiology due to its high-contrast resolution, fast imaging speed, and many utilities. With the increasing popularity of CT scans, there is a major safety concern on ionizing radiation from x-ray exposure to the patient. To reduce the x-ray radiation-induced risks during a CT scan, low-dose CT (LDCT) research has attracted a widespread attention so that the image quality of LDCT can be optimized despite the radiation dose reduction. 
	
	To solve the above challenge of LDCT, there have been numerous efforts made in the past few decades. A natural approach for LDCT is to post-process a noisy LDCT image. 
	Inspired by established methods for natural image denoising, non-local means (NLM) and block-matching 3D (BM3D) were adapted for LDCT denoising~\cite{li2014adaptive, feruglio2010block, kang2013image}, significantly improving the LDCT performance. Although these post-processing methods do remove LDCT image noise, the resultant image quality often hardly meets the clinical requirements, since it is difficult to model the noise contained in the LDCT images, and thus new artifacts can be frequently introduced. Another commonly used LDCT denoising approach is referred to as model-based iterative reconstruction (MBIR), which is to establish a data model in the projection domain and a content model in the image domain, and then minimize an objective function for model fitting iteratively. Especially, inspired by compressed sensing (CS) theory, the regularization model with total variation (TV) greatly improves the reconstruction quality~\cite{yu2005total}. Subsequently, many methods were proposed to solve the over-smoothness problem associated with TV minimization, including anisotropic TV~\cite{chen2013limited}, total generalized variation (TGV)~\cite{niu2014sparse}, fractional order TV~\cite{zhang2014few, zhang2016statistical}, and so on. In addition to the TV regularization, other effective priors were also formulated, such as NLM~\cite{chen2009bayesian}, tight framework~\cite{gao2011multi}, low-rank~\cite{cai2014cine}, dictionary learning~\cite{xu2012low, chen2013improving},
	learned sparsifying transform~\cite{zheng2016low}, etc. These regularization-based methods can achieve a promising denoising performance by careful designing the regularization term and fine-tuning the hyperparameters. However, that means extensive experience and skills, and limited clinical applications.
	
	With the recent development of the deep learning (DL) techniques~\cite{lecun2015deep}, LDCT denoising was immediately identified as an earliest application target~\cite{wang2016perspective}. Chen \textit{et al.} first published a convolutional neural network (CNN) for LDCT denoising and achieved remarkable performance~\cite{chen2017lowdose}. Progress in this area has been rapid, including residual structures to improve both the training convergence and denoising performance~\cite{chen2017low, jin2017deep, kang2017deep, han2018framing}, and incorporate various loss functions and training strategies, such as adversarial loss~\cite{wolterink2017generative, yang2018low}, perceptual loss~\cite{yang2018low}, edge incoherence loss~\cite{shan2019competitive}, etc. Also, in some studies DL and MBIR algorithms were combined for synergies~\cite{wu2017iterative, chen2018learn, adler2018learned, gupta2018cnn, chun2020momentum, xiang2021fista}. These methods deploy CNNs in iterative optimization schemes so that the priors and penalty parameters can be learned in a data-driven fashion.
	
	Very recently, the denoising diffusion probabilistic model (DDPM)~\cite{sohl2015deep, ho2020denoising, croitoru2022diffusion, yang2022diffusion} emerged as a generative model with the mode diversity and output quality superior to that of the generative adversarial network (GAN)~\cite{goodfellow2020generative,adler2018learned, sohl2015deep, ho2020denoising, song2019generative, song2020score, dhariwal2021diffusion, nichol2021improved}.
	DDPM gradually perturbs an original data distribution until into a standard distribution, typically a normal distribution. Then, DDPM iteratively recovers the data distribution in a learned reverse diffusion process. Song \textit{et al.} interpreted the forward and backward diffusion processes in the stochastic differential equation (SDE) framework~\cite{song2020score}. The SDE-based diffusion model allows the learning of the reverse diffusion process in continuous time, and has been applied to a variety of tasks, such as image super-resolution~\cite{rombach2022high, saharia2022image}, image inpainting~\cite{lugmayr2022repaint}, image editing~\cite{meng2021sdedit}, image translation~\cite{choi2021ilvr, saharia2022palette}, and so on.
	
	In this paper, we introduce a conditional DDPM for LDCT denoising. The U-Net~\cite{ho2020denoising} is adopted to learn the reverse diffusion process of a normal-dose CT (NDCT) image conditioned on a LDCT counterpart. Then, by gradually sampling, cleaner images can be step by step recovered from a normal noise distribution under the condition of the LDCT image. The final image generated by DDPM can remove structural noise in the input LDCT image while preserving clinically important details. However, the sampling efficiency of the original DDPM model is low and can hardly meet clinical requirements. To alleviate this problem, here we use a fast ordinary differential equation (ODE) solver~\cite{lu2022dpm} for a much-boosted efficient sampling, 20 $\times$ faster than that of the original DDPM. 
	 
\section{Methodology}

\subsection{Supervised Deep LDCT Denoising}
	Currently, a typical DL-based LDCT denoising method trains a network to learn a mapping from LDCT images to NDCT images. Let us assume that $ \bm{x} \in \mathbb{R}^{N} $ and $ \bm{y} \in \mathbb{R}^N $ are paired LDCT and NDCT images, then the parameters of the network can be trained as follows:
	\begin{equation}
		\min_{\bm{\theta}} \left\| D_{\bm{\theta}}(\bm{x}) - \bm{y} \right\|_2^2,
		\label{eq:1}
	\end{equation}
	where $ D_{\bm{\theta}}: \mathbb{R}^N \rightarrow \mathbb{R}^N $ is the network defined by $ \bm{\theta} $, which is a vector of parameters.
	
	\begin{figure*}[tbp]
		\centering
		\includegraphics[width=1.0\textwidth]{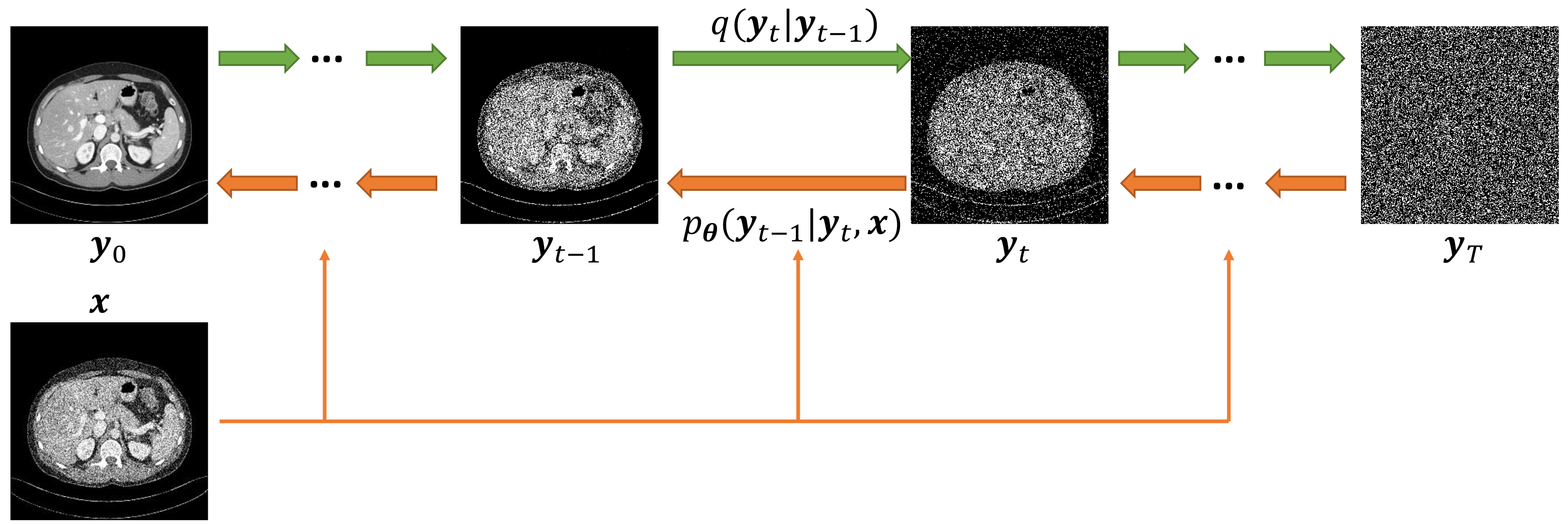}
		\caption{Conditional DDPM for LDCT denoising, trained in the supervised mode.}
		\label{fig:1}
	\end{figure*}
	
\subsection{Conditional DDPM for LDCT Denoising}
	As shown in Fig.~\ref{fig:1}, the conditional DDPM for LDCT denoising consists of a \textit{forward process} and a \textit{reverse process}. In reference to~\cite{ho2020denoising}, the \textit{forward process} is defined as a Markov chain which gradually adds Gaussian noise to an NDCT image $ \bm{y}_0 $ according to a hyper-parameterized variance sequence $\beta_1, \beta_2, \cdots, \beta_T  $:
	\begin{equation}
		q(\bm{y}_{1:T}|\bm{y}_{0}) = \prod_{t=1}^T q(\bm{y}_t|\bm{y}_{t-1}),
		\label{eq:2}
	\end{equation}
	where
	\begin{equation}
		q(\bm{y}_t|\bm{y}_{t-1}) = \mathcal{N} (\bm{y}_t| \sqrt{1 - \beta_t}\bm{y}_{t-1}, \beta_t \bm{I}).
		\label{eq:3}
	\end{equation}
	
	According to the properties of the Gaussian distribution, the sampling operation can be directly performed at an arbitrary timestep $ t $ without any iteration:
	\begin{equation}
		q(\bm{y}_t|\bm{y}_{0}) = \mathcal{N} (\bm{y}_t| \sqrt{\bar{\alpha}_t}\bm{y}_0, 	(1-\bar{\alpha}_t) \bm{I}),
		\label{eq:4}
	\end{equation}
	where $\alpha_t = 1-\beta_t$ and $ \bar{\alpha}_t = \prod_{i=1}^t \alpha_i $. After the forward process, $ \bm{y}_T $ will follow a standard normal distribution $ \mathcal{N}(0,\bm{I}) $ when $ T $ is large enough. Then, it can be shown that the posterior distribution of $ \bm{y}_{t-1} $ given $ \bm{y}_{t} $ and $ \bm{y}_0 $  can be formulated as follows:
	\begin{equation}
		\left\lbrace 
		\begin{aligned}
			&q(\bm{y}_{t-1}|\bm{y}_{t}, \bm{y}_0) = 	\mathcal{N}(\bm{y}_{t-1}|\tilde{\mu}_t(\bm{y}_{t}, \bm{y}_{0}), \sigma_t^2 \bm{I}),\\
			&\tilde{\mu}_t(\bm{y}_{t}, \bm{y}_{0}) = 	\frac{\sqrt{\alpha_t}(1-\bar{\alpha}_{t-1})}{1- \bar{\alpha}_t} \bm{y}_t + \frac{\sqrt{\bar{\alpha}_{t-1}}(1-\alpha_t)}{1- \bar{\alpha}_t} \bm{y}_0, \\
			&\sigma_t^2 = \frac{(1-\bar{\alpha}_{t-1})(1-\alpha_t)}{1- \bar{\alpha}_t}.
		\end{aligned}
		\right.	
		\label{eq:5}
	\end{equation}
	
	Similar to the forward process, the \textit{reverse process} conditioned on a LDCT image $ \bm{x} $ is also a Markov chain defined by
	\begin{equation}
		p_{\theta} (\bm{y}_{0:T}|\bm{x})=p(\bm{y}_{T})\prod_{t=1}^T 	p_{\theta}(\bm{y}_{t-1}|\bm{y}_{t}, \bm{x}),
		\label{eq:6}
	\end{equation}
	where
	\begin{equation}
		p_{\theta}(\bm{y}_{t-1}|\bm{y}_{t}, \bm{x}) = 	\mathcal{N}(\bm{y}_{t-1}|\mu_{\theta}(\bm{y}_{t}, \bm{x}, t), \sigma_t^2 \bm{I}),
		\label{eq:7}
	\end{equation}
	where $ \bm{y}_{T} $ is sampled from a standard normal distribution $ p(\bm{y}_{T}) \sim \mathcal{N}(0, \bm{I}) $, $ \mu_{\theta} $ is the expectation computed by a learned U-Net model~\cite{ho2020denoising}.
	
	The neural network is trained with a variational lower bound (VLB) of the negative log-likelihood:
	\begin{equation}
		\mathcal{L}_{VLB} = \mathbb{E}_{q} \left[ 	-\log\frac{p_{\theta}(\bm{y}_{0:T}|\bm{x})}{q(\bm{y}_{1:T}|\bm{y}_0)} \right]\geq \mathbb{E} \left[-\log p_{\theta} (\bm{y}_0|\bm{x})\right].
		\label{eq:8}
	\end{equation}
	
	After inserting the Gaussian density function into Eq. (\ref{eq:8}) and removing the constant, the objective function can be rewritten as
	\begin{equation}
		\mathcal{L} = \mathbb{E}_{\bm{x},\bm{y}}\mathbb{E}_{t} \left[\frac{1}{2\sigma_t^2} \left\| 	\tilde{\mu}_t(\bm{y}_{t}, \bm{y}_{0}) - \mu_{\theta}(\bm{y}_{t}, \bm{x}, t) \right\|_2^2 \right].
		\label{eq:9}
	\end{equation}
	
	Specifically, the forward process can be implemented as
	\begin{equation}
		\bm{y}_t = \sqrt{\bar{\alpha}_t}\bm{y}_0 + \sqrt{1-\bar{\alpha}_t} \bm{\epsilon}, \ 	\bm{\epsilon}\sim \mathcal{N}(0, \bm{I}),
		\label{eq:10}
	\end{equation}
	Then, the posterior expectation in Eq. (\ref{eq:5}) can be rewritten as
	\begin{equation}	
		\tilde{\mu}_t(\bm{y}_{t}, \bm{y}_{0}) = \tilde{\mu}_t\left(\bm{y}_{t}, 	\frac{1}{\sqrt{\bar{\alpha}}}\left(\bm{y}_t-\sqrt{1-\bar{\alpha}}\bm{\epsilon}\right)\right)
		=\frac{1}{\sqrt{\bar{\alpha}}}\left(\bm{y}_t-\frac{1-\alpha_t}{\sqrt{1-\bar{\alpha}}}\bm{\epsilon}\right).
		\label{eq:11}
	\end{equation}
	
	Instead of directly learning the expectation, the U-Net model $ D_{\bm{\theta}} $ is trained to predict the noise vector $ \bm{\epsilon} $. Then, the objective function Eq. (\ref{eq:9}) becomes
	\begin{equation}
		\mathcal{L} = \mathbb{E}_{\bm{x},\bm{y}}\mathbb{E}_{\bm{\epsilon},t} 	\left[\frac{(1-\alpha_t)^2}{2\sigma_t^2 \bar{\alpha}(1-\bar{\alpha})} \left\| \bm{\epsilon} - D_{\bm{\theta}} (\sqrt{\bar{\alpha}_t}\bm{y}_0 + \sqrt{1-\bar{\alpha}_t} \bm{\epsilon}, \bm{x}, t) \right\|_2^2 \right].
		\label{eq:12}
	\end{equation}
	According to~\cite{ho2020denoising}, we remove the weight of the $\ell_2$ norm for better training performance. Finally, for inference the sampling operation $ \bm{y}_{t-1} \sim p_{\theta}(\bm{y}_{t-1}|\bm{y}_{t})$ can be computed as follows:
	\begin{equation}
		\bm{y}_{t-1} = 	\frac{1}{\sqrt{\bar{\alpha}}}\left(\bm{y}_t-\frac{1-\alpha_t}{\sqrt{1-\bar{\alpha}}}D_{\bm{\theta}} (\bm{y}_t, \bm{x}, t)\right) +\sigma_t \bm{z}, \ \bm{z}\sim \mathcal{N}(0,\bm{I}).
	\end{equation}
	For clarify, Algorithms \ref{alg:1} and \ref{alg:2} are presented in the pseudo codes for training and inference, respectively.
	
	\begin{algorithm}[t]
		\caption{Training of the denoising model $ D_{\bm{\theta}} $.}
		\label{alg:1}
		\KwIn{Number of time steps $T$; Variance schedule $\{\beta_t|t=1,2,...,T\}$}
		\KwOut{Trained model $ D_{\bm{\theta}} $}
		\BlankLine
		Initialize $D_{\bm{\theta}}$ randomly;
		
		\While{\textnormal{not converged}}{
			$ (\bm{x}, \bm{y}_0) \sim p(\bm{x}, \bm{y}) $
			
			$ t\sim \mathrm{Uniform}(\{1,2,...,T\}) $
			
			$ \bm{\epsilon} \sim \mathcal{N}(0, \bm{I}) $
			
			Update $\bm{\theta}$ with the gradient $\nabla_{\theta} \left\| \bm{\epsilon} - D_{\bm{\theta}} (\sqrt{\bar{\alpha}_t}\bm{y}_0 + \sqrt{1-\bar{\alpha}_t} \bm{\epsilon}, \bm{x}, t) \right\|_2^2$
		}
	\end{algorithm}
	
	\begin{algorithm}[t]
		\caption{Inference with the trained denoising model $ D_{\bm{\theta}} $.}
		\label{alg:2}
		\KwIn{Number of time steps $T$; Variance schedule $\{\beta_t|t=1,2,...,T\}$}
		\KwOut{$\bm{y}_0$}
		\BlankLine
		
		Load $ D_{\bm{\theta}} $;
		
		$ \bm{y}_T \sim \mathcal{N}(0, \bm{I})$
		
		\For{$t=1,2,...,T$}{
		$\bm{z} \sim \mathcal{N}(0, \bm{I})$ if $t>1$ else $\bm{z}=0$
		
		$\bm{y}_{t-1} = \frac{1}{\sqrt{\bar{\alpha}}}\left(\bm{y}_t-\frac{1-\alpha_t}{\sqrt{1-\bar{\alpha}}}D_{\bm{\theta}} (\bm{y}_t, \bm{x}, t)\right) +\sigma_t \bm{z}$
		}
		
	\end{algorithm}

\subsection{Fast ODE Solver}
	Because the inference phase originally requires thousands of sampling steps, the computational cost is too expensive to meet clinical requirements. Therefore, it is necessary to accelerate the sampling process of DDPM. In~\cite{song2020score}, Song \textit{et al.} proved that the reverse process of DDPM has a form equivalent to a \textit{probability flow ODE}. Thus, the RK45 ODE solver~\cite{dormand1980family} can be applied to greatly reduce the number of sampling steps. In this study, we introduce a fast ODE solver to further improve the sampling efficiency.
	
	To cast the discrete diffusion processes of DDPM in the continuous form, Kingma \textit{et al.}~\cite{kingma2021variational} proved that the forward process for $ t\in\{1,2,...,T\} $ following the distribution Eq. (\ref{eq:4}) is equivalent to a stochastic differential equation (SDE) starting from $ \bm{y}_0 \sim p(\bm{y}_0) $ for any $ t\in[0,T] $:
	\begin{equation}
		\mathrm{d}\bm{y} = f(t) \bm{y} \mathrm{d}t + g(t)\mathrm{d}\bm{w}, 
		\label{eq:14}
	\end{equation}
	where $ \bm{w}_t \in \mathbb{R}^N $ is the standard Wiener process, and	
	\begin{equation}
		f(t) = \frac{\mathrm{d} \log{\eta_t}}{\mathrm{d}t}, \qquad g^2(t)=\frac{\mathrm{d}\sigma_t^2}{\mathrm{d} t} - 2\frac{\mathrm{d} \log{\eta_t}}{\mathrm{d}t}\sigma_t^2,
		\label{eq:14.1}
	\end{equation} 
	where $ \eta_t =  \sqrt{\bar{\alpha}_t}$. Starting from $ \bm{y}_T \sim p(\bm{y}_T) $, the reverse process can be formulated as a reverse-time SDE~\cite{song2020score}:
	\begin{equation}
		\mathrm{d}\bm{y} = \left[f(t) \bm{y} - g^2(t) \nabla_{\bm{y}} \log{q_t(\bm{y})} \right]\mathrm{d}t + g(t)\mathrm{d}\bar{\bm{w}}, 
		\label{eq:15}
	\end{equation}
	where $ \bar{\bm{w}} \in \mathbb{R}^N$ is the standard Wiener process for the reverse time from $ T $ to $ 0 $, $ \log q_t(\bm{y}) $ is called the \textit{score function} and can be learned with a neural network~\cite{song2020score}. The reverse process of DDPM can be regarded as a specific form of Eq. (\ref{eq:15}), in which the noise prediction model $ D_{\bm{\theta}}(\bm{y}_t, \bm{x},t) $ is to learn $ -\sigma_t \nabla_{\bm{y}} \log{q_t(\bm{y})} $. The sampling of the above SDE with a large step will hardly converge because of the randomness of the Wiener process~\cite{platen2010numerical}. In~\cite{song2020score}, Song \textit{et al.} proved that the SDE shown in Eq. (\ref{eq:15}) has the same marginal distribution at each time $ t $ as that of a probability flow ODE, which can be sampled with a larger step:
	\begin{equation}
		\mathrm{d} \bm{y} = \left[ f(t) \bm{y} - \frac{1}{2} g^2(t) \nabla_{\bm{y}} \log{q_t(\bm{y})} \right] \mathrm{d} t,
		\label{eq:16}
	\end{equation}
	Substituting the noise prediction model into Eq. (\ref{eq:16}), the reverse process can be implemented as
	\begin{equation}
		\mathrm{d} \bm{y} = \left[ f(t) \bm{y} + \frac{g^2(t)}{2\sigma_t}  D_{\bm{\theta}}(\bm{y}_t, \bm{x},t) \right] \mathrm{d} t,
		\label{eq:17}
	\end{equation}
	which is a semi-linear ODE~\cite{lu2022dpm}, whose solution at time $ t $ can be obtained with the \textit{variation of constants} formula~\cite{atkinson2011numerical}:
	\begin{equation}
		\bm{y}_t = \exp\left(\int_{s}^{t}f(\tau) \mathrm{d} \tau\right) \bm{y}_s+ \int_{s}^{t}\left[ \exp\left(\int_{\tau}^{t}f(r) \mathrm{d}r\right) \frac{g^2(\tau)}{2\sigma_t} D_{\bm{\theta}}(\bm{y}_{\tau}, \bm{x}, \tau) \right]\mathrm{d}\tau.
		\label{eq:18}
	\end{equation}
	Let $ \lambda_t = \log(\eta_t/\sigma_t) $, Eq. (\ref{eq:18}) can be further simplified into
	\begin{equation}
		\bm{y}_t = \frac{\eta_t}{\eta_s} \bm{y}_s + \eta_t \int_s^t \left[\left(\frac{\mathrm{d}\lambda_{\tau}}{\mathrm{d}\tau}\right)\frac{\sigma_{\tau}}{\eta_{\tau}}D_{\theta}(\bm{y}_{\tau}, \bm{x}, \tau)\right]\mathrm{d}\tau.
		\label{eq:19}
	\end{equation}
	The predefined $ \lambda_t $ can be obtained with a strictly decreasing function of $ t $, denoted as $ \lambda(t) $, which has a inverse function $ t = t_{\lambda}(\lambda) $. Then, by changing the time variable $t$ into the parameter variable $\lambda$ and denoting $ \hat{\bm{y}}_{\lambda}:= \bm{y}_{t_{\lambda}(\lambda)}$ and $ \hat{D}_{\bm{\theta}} (\hat{\bm{y}}_{\lambda}, \bm{x}, \lambda):= D_{\bm{\theta}}(\bm{y}_{t_{\lambda}(\lambda)}, \bm{x}, t_{\lambda}(\lambda))$, Eq. (\ref{eq:19}) can be rewritten as
	\begin{equation}
		\bm{y}_t = \frac{\eta_t}{\eta_s} \bm{y}_s + \eta_t \int_{\lambda_{s}}^{\lambda{t}} \left[\mathrm{e}^{-\lambda} \hat{D}_{\bm{\theta}} (\hat{\bm{y}}_{\lambda}, \bm{x}, \lambda)\right] \mathrm{d} \lambda,
		\label{eq:20}
	\end{equation}
	where the integral $ \int \mathrm{e}^{-\lambda} \hat{D}_{\bm{\theta}} \mathrm{d} \lambda$ is called the \textit{exponentially weighted integral} of $ \hat{D}_{\bm{\theta}} $~\cite{lu2022dpm}. This integral can be calculated numerically by Taylor expansion. According to the order of Taylor expansion, Lu \textit{et al.}~\cite{lu2022dpm} provided three solvers for the flow probability ODE. Algorithms \ref{alg:3}, \ref{alg:4} and \ref{alg:5} are the ODE solvers obtained from the first-order, second-order and third-order Taylor expansion, which are called \textit{DPM-Solver-1}, \textit{DPM-Solver-2} and \textit{DPM-Solver-3}, respectively. As shown in the pseudo codes of the algorithms, there are $k$ times of functional evaluation in DPM-Solver-$k$ per step. The higher order solver has a faster convergence speed so that it takes fewer steps to achieve satisfactory results~\cite{lu2022dpm}. Therefore, with a limited number of functional evaluations (NFE), the DPM-Solver-3 is recommended. After applying DPM-Solver-3, if the remaining NFEs can no longer be divisible by 3, we can apply DPM-Solver-1 or DPM-Solver-2.
	
	\begin{algorithm}[t]
		\caption{DPM-Solver-1.}
		\label{alg:3}
		\KwIn{Time sequence $\{t_i\in[0,1]|i=0,1,...,M\}$} 
		\KwOut{$\tilde{\bm{y}}_{t_M}$}
		\BlankLine
		
		Load $ D_{\bm{\theta}} $
		
		$ \bm{y}_T \sim \mathcal{N}(0, \bm{I})$
		
		$\tilde{\bm{y}}_{t_0} = \bm{y}_T$
		
		\For{$i=1,2,...,M$}{
			$h_i = \lambda_{t_i}-\lambda_{t_{i-1}}$
			
			$ \tilde{\bm{y}}_{t_i} = \frac{\eta_{t_i}}{\eta_{t_{i-1}}} \tilde{\bm{y}}_{t_{i-1}} - \sigma_{t_i}\left(\mathrm{e}^{h_i}-1\right) D_{\bm{\theta}}(\tilde{\bm{y}}_{t_{i-1}}, \bm{x}, t_{i-1})$
		}
		
	\end{algorithm}
	
	\begin{algorithm}[t]
		\caption{DPM-Solver-2.}
		\label{alg:4}
		\KwIn{Time sequence $\{t_i\in[0,1]|i=0,1,...,M\}$} 
		\KwOut{$\tilde{\bm{y}}_{t_M}$}
		\BlankLine
		
		Load $ D_{\bm{\theta}} $
		
		$ \bm{y}_T \sim \mathcal{N}(0, \bm{I})$
		
		$\tilde{\bm{y}}_{t_0} = \bm{y}_T$
		
		\For{$i=1,2,...,M$}{
			$h_i = \lambda_{t_i}-\lambda_{t_{i-1}}$
			
			$s_i = t_{\lambda} (\frac{\lambda_{t_{i-1}}+\lambda_{t_i}}{2})$
			
			$ \bm{u}_{i} = \frac{\eta_{s_i}}{\eta_{t_{i-1}}} \tilde{\bm{y}}_{t_{i-1}} - \sigma_{s_i}\left(\mathrm{e}^\frac{{h_i}}{2}-1\right) D_{\bm{\theta}}(\tilde{\bm{y}}_{t_{i-1}}, \bm{x}, t_{i-1})$
			
			$ \tilde{\bm{y}}_{t_i} = \frac{\eta_{t_i}}{\eta_{t_{i-1}}} \tilde{\bm{y}}_{t_{i-1}} - \sigma_{t_i}\left(\mathrm{e}^{h_i}-1\right) D_{\bm{\theta}}(\bm{u}_{i}, \bm{x}, s_i)$
		}
		
	\end{algorithm}

	\begin{algorithm}[t]
		\caption{DPM-Solver-3.}
		\label{alg:5}
		\KwIn{Time sequence $\{t_i\in[0,1]|i=0,1,...,M\}$} 
		\KwOut{$\tilde{\bm{y}}_{t_M}$}
		\BlankLine
		
		Load $ D_{\bm{\theta}} $
		
		$ \bm{y}_T \sim \mathcal{N}(0, \bm{I})$
		
		$\tilde{\bm{y}}_{t_0} = \bm{y}_T,\ r_1=\frac{1}{3},\ r_2=\frac{2}{3}$
		
		\For{$i=1,2,...,M$}{
			$h_i = \lambda_{t_i}-\lambda_{t_{i-1}}$
			
			$s_{2i-1} = t_{\lambda} (\lambda_{t_{i-1}}+r_1h_i),\ s_{2i} = t_{\lambda} (\lambda_{t_{i-1}}+r_2h_i)$
			
			$ \bm{u}_{2i-i} = \frac{\eta_{s_{2i-1}}}{\eta_{t_{i-1}}} \tilde{\bm{y}}_{t_{i-1}} - \sigma_{s_{2i-1}}\left(\mathrm{e}^{r_1 h_i}-1\right) D_{\bm{\theta}}(\tilde{\bm{y}}_{t_{i-1}}, \bm{x}, t_{i-1})$
			
			$ \bm{v}_{2i-1} = D_{\bm{\theta}}(\bm{u}_{2i-1}, \bm{x}, s_{2i-1}) -  D_{\bm{\theta}}(\tilde{\bm{y}}_{t_{i-1}}, \bm{x}, t_{i-1})$
			
			$ \bm{u}_{2i} = \frac{\eta_{s_{2i}}}{\eta_{t_{i-1}}} \tilde{\bm{y}}_{t_{i-1}} - \sigma_{s_{2i}}\left(\mathrm{e}^{r_2 h_i}-1\right) D_{\bm{\theta}}(\tilde{\bm{y}}_{t_{i-1}}, \bm{x}, t_{i-1}) - \frac{\sigma_{s_{2i}} r_2}{r_1}\left(\frac{\mathrm{e}^{r_2 h_i}-1}{r_2h_i}-1\right) \bm{v}_{2i-1}$
			
			$ \bm{v}_{2i-1} = D_{\bm{\theta}}(\bm{u}_{2i}, \bm{x}, s_{2i}) -  D_{\bm{\theta}}(\tilde{\bm{y}}_{t_{i-1}}, \bm{x}, t_{i-1})$
			
			$ \tilde{\bm{y}}_{t_i} = \frac{\eta_{t_i}}{\eta_{t_{i-1}}} \tilde{\bm{y}}_{t_{i-1}} - \sigma_{t_i}\left(\mathrm{e}^{h_i}-1\right) D_{\bm{\theta}}(\tilde{\bm{y}}_{t_{i-1}}, \bm{x}, t_{i-1}) - \frac{\sigma_{t_i}}{r_2}\left(\frac{\mathrm{e}^{h_i} - 1}{h_i} - 1\right) \bm{v}_{2i}$
		}
		
	\end{algorithm}

	The DDPM model is trained with discrete time steps, while the DPM-Solver solves the SDE in a continuous fashion. Nevertheless, we can still apply the DPM-Solver to an already trained DDPM model. When applying the DPM-solver, we discretize the continuous time sequence into DDPM steps:	
	\begin{equation}
		t_d = 1000 \cdot \max\left( t_c - \frac{1}{T}\right),
		\label{eq:21}
	\end{equation}
	where $t_d$ and $t_c$ are discrete and continuous time variables respectively, and $T$ is the total number of discrete time steps.

	\begin{figure*}[tbp]
		\centering
		\includegraphics[width=0.8\textwidth]{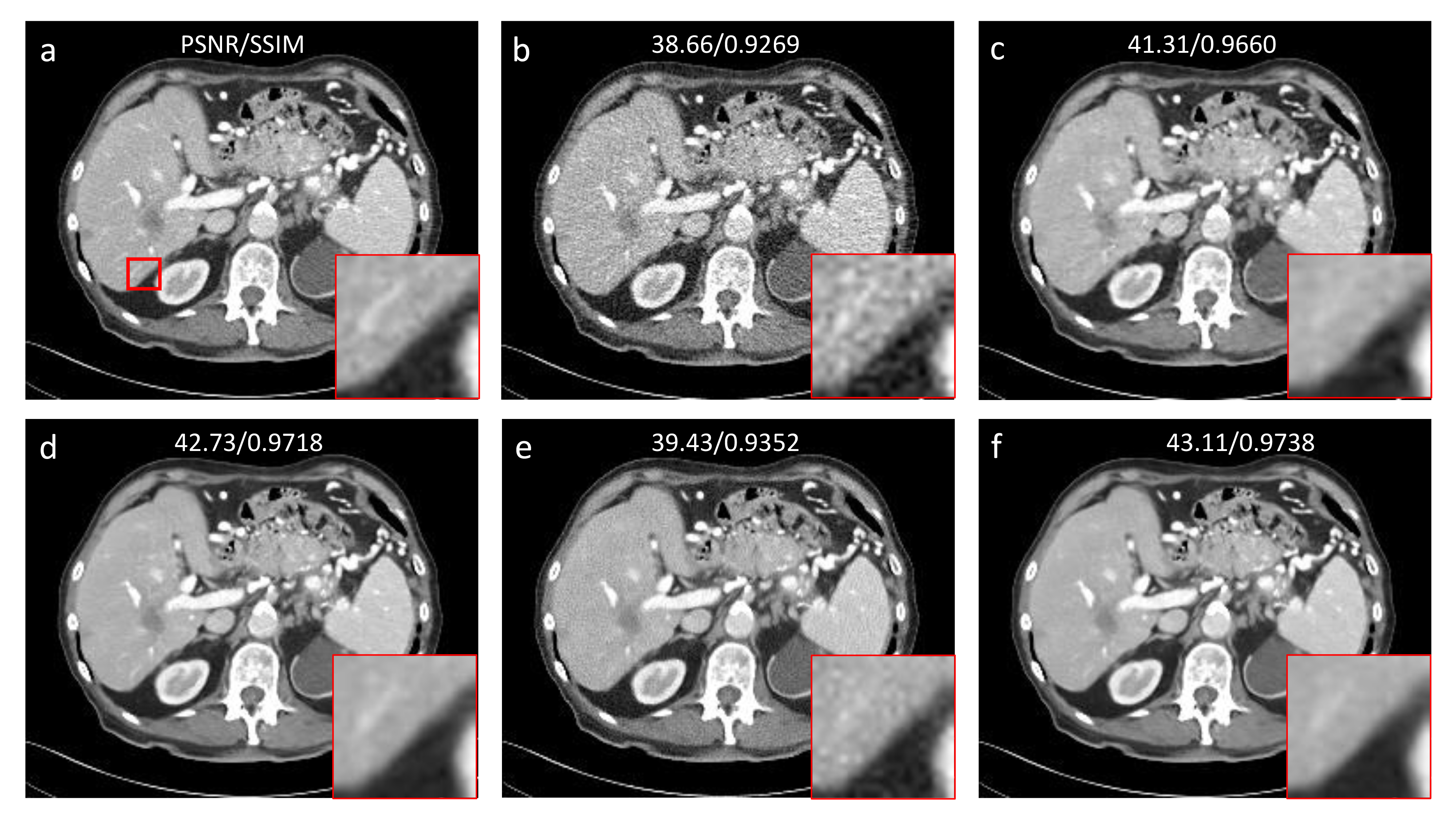}
		\caption{The results obtained using different methods. (a) NDCT, (b) LDCT, (c) U-Net, (d) DDPM, (e) DPM-Solver 15 NFE, and (f) DPM-Solver 50 NFE. The display window is consistently set to [-160, 240] HU.}
		\label{fig:2}
	\end{figure*}

\section{Experiments and Results}
	To evaluate the performance of our proposed method for LDCT denoising, the \textit{2016 NIH-AAPM-Mayo Clinic Low-Dose CT Grand Challenge} dataset was chosen. The dataset has 2,378 paired NDCT and LDCT images with a thickness of 3mm from 10 patients. We selected 1,923 paired images from 8 patients as the training set, and 455 paired images from the remaining 2 patients as the test set. The image matrix was resampled to be 256x256.
	The total number of time steps $T$ was set to 1,000. The model was trained using the Adam optimizer~\cite{kingma2014adam} with a learning rate of $ 1\times 10^{-4} $. The training process converged well after $ 5\times10^{5} $ iterations on a computing server equipped with Nvidia GTX 1080 Ti x4 GPUs. After the model was trained, we used DDPM, DPM-Solver 15 NFE and DPM-Solver 50 NFE to sample the test set respectively.
	
\begin{figure*}[tbp]
		\centering
		\includegraphics[width=0.8\textwidth]{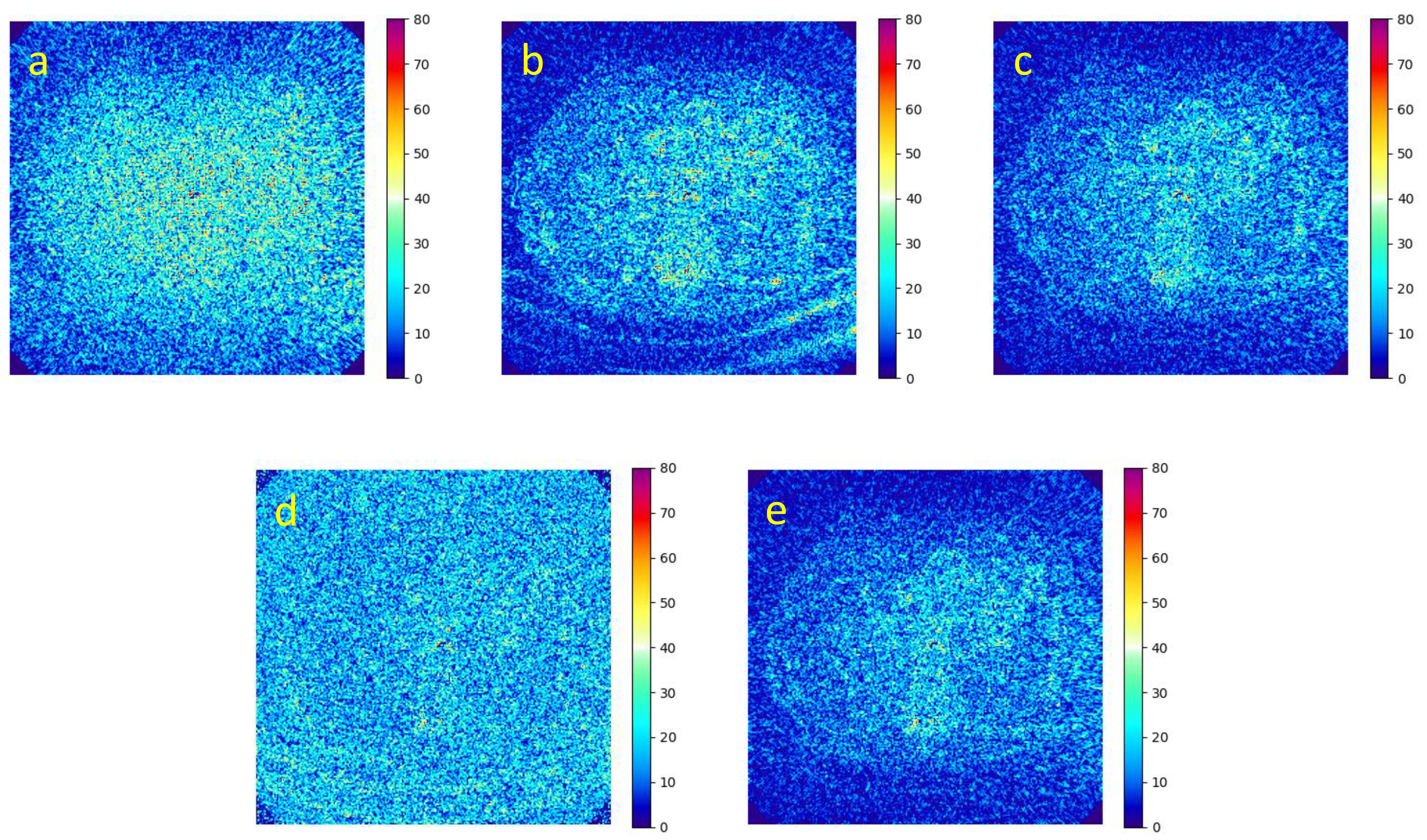}
		\caption{The absolute error maps of the results in reference to the NDCT image. (a) LDCT, (b) U-Net, (c) DDPM, (d) DPM-Solver 15 NFE, and (e) DPM-Solver 50 NFE. The display window is [0, 80] HU.}
		\label{fig:3}
	\end{figure*}

	Fig.~\ref{fig:2} shows the results obtained using different methods. It can be seen that U-Net~\cite{ronneberger2015u} failed to eliminate image noise completely. With DDPM, image noise was effectively removed while the structures were well kept. With only 15 NFE, the DPM-Solver could not denoise LDCT images well. Because of a too few number of sampling steps, residual noise and newly generated noise are evident in the resultant images sampled with DPM-Solver 15 NFE. Remarkably, after NFE was increased to 50, the DPM-solver delivered a competitive performance relative to that of DDPM. It can be seen that in the magnified region of interest (ROI), DPM-Solver 50 NFE well maintained the vasculature, which was blurred by U-Net. Quantitatively, DPM-Solver 50 NFE  outperformed DDPM. Fig.~\ref{fig:3} shows the absolute error maps of the abdominal results shown in Fig.~\ref{fig:2} in reference to the NDCT image. It further demonstrates the efficient and effective sampling performance of the DPM-Solver, whose error map is the least noisy.

	Table~\ref{tab:1} shows the averaged quantitative evaluation of different methods over the whole test set. It can be seen that DDPM and DPM-Solver 50 NFE have the best quantitative scores. Especially, DPM-Solver 50 NFE quantitatively outperforms DDPM, which means that sampling with the DPM-Solver gives a superior noise suppression performance.
	
	\begin{table}[t]
		\centering
		\caption{Average quantitative results of different methods.}
		\label{tab:1}
		\begin{tabular}{ccc}
			\toprule
			Method            & PSNR  & SSIM   \\
			\midrule
			LDCT              & 42.81 & 0.9667 \\
			U-Net             & 44.08 & 0.9823 \\
			DDPM              & 46.49 & 0.9873 \\
			DPM-Solver 15 NFE & 40.41 & 0.9423 \\
			DPM-Solver 50 NFE & 46.72 & 0.9880 \\
			\bottomrule
		\end{tabular}
	\end{table}

	Table~\ref{tab:2} shows the averaged computational costs of different methods. Without any need for iterative sampling, U-Net has the highest processing efficiency. However, the denoising performance of U-Net is of limited value in clinical applications. DDPM requires a very high inference cost, which is hardly acceptable in practice. The DPM-Solver allows an excellent acceleration, while keeping or even improving the sampling quality. It can be seen that DPM-Solver 50 NFE achieves a 20$\times$ speedup compared to DDPM while maintaining denoising performance.

	\begin{table}[t]
		\centering
		\caption{Averaged inference costs of different methods}
		\label{tab:2}
		\begin{tabular}{cc}
			\toprule
			Method             &  Cost (second per image)  \\
			\midrule
			U-Net              &  0.08  \\
			DDPM               &  62.17 \\
			DPM-Solver 15 NFE  &  0.95 \\
			DPM-Solver 50 NFE  &  3.12 \\                 
			\bottomrule              
		\end{tabular}
	\end{table}

\section{Conclusion}
	In this paper, we have adapted a novel conditional DDPM model for LDCT denoising. The denoising performance outperforms the traditional CNN-based method. Furthermore, to accelerate the sampling operation of DDPM, we have applied a fast ODE solver. It is shown that the ODE Solver has both a high denoising performance and a high sampling efficiency. Specifically, the ODE solver can achieve a 20$\times$ speedup without compromising the denoising performance. The superior performance of the ODE solver for DDPM sampling means a great potential for clinical applications.

\bibliographystyle{plain} 
\bibliography{reference}

\end{document}